\documentclass{article}%%\documentstyle{article}

\usepackage{graphicx,latexsym}
\input epsf.sty

\def\abstract#1{\vskip 7mm 
        \begin{center}{\large Abstract}\par \smallskip
                \begin{minipage}[c]{12cm}
                        \small #1
                \end{minipage}
        \end{center}
}
\def\title#1{\begin{center}{\Large\bf #1}\end{center}}\def\author#1{\vskip 5mm \begin{center}{#1}\end{center}}
\def\address#1{\begin{center}{\it #1}\end{center}}
\makeatletter
\@ifundefined{lesssim}{}{}
\@ifundefined{gtrsim}{}{}
\def\vereq#1#2{\lower3pt\vbox{\baselineskip1.5pt \lineskip1.5pt
\ialign{$\m@th#1\hfill##\hfil$\crcr#2\crcr\sim\crcr}}}
\makeatother

\begin{document}

\title{%
  New method to integrate 2+1 wave equations\\ with Dirac's delta functions as sources
  \smallskip \\
%  {\large --- Please use this file to complete your manuscript  ---}
}
\author{%
  Hiroyuki Nakano\footnote{E-mail:hxnsma1@rit.edu}
  and
  Carlos O. Lousto\footnote{E-mail:colsma@rit.edu}
}
\address{%
  $^{1,2}$Center for Computational Relativity and Gravitation, 
School of Mathematical Sciences,\\
Rochester Institute of Technology, Rochester, New York 14623, USA
}

\abstract{
Gravitational perturbations in a Kerr black hole background 
can not be decomposed into simple tensor harmonics in the time domain. 
Here, we make the mode decomposition only in the azimuthal direction 
and discuss the resulting (2+1)-dimensional Klein-Gordon differential equation 
for scalar perturbations with a two dimensional Dirac's $\delta$-function 
as a source representing a point particle orbiting a much larger black hole. 
To make this equation amenable for numerical integrations we 
explicitly remove analytically the singular behavior of the source 
and compute a global effective source for the corresponding waveform.
}

%%%%%%%%%%%%%%%%%%%%%%%%%%%%%%%%%%%%%%%%%%%%%%%%%%%%%%%%%%%%%%%%%%%%%%
\section{Introduction}
%%%%%%%%%%%%%%%%%%%%%%%%%%%%%%%%%%%%%%%%%%%%%%%%%%%%%%%%%%%%%%%%%%%%%%

One of the main source targets of LISA is the gravitational waves 
generated by the inspiral of compact objects 
into massive black holes. For these Extreme Mass Ratio Inspirals (EMRI) 
we use the black hole perturbation approach to compute waveforms, 
where the compact object is approximated by a point particle 
orbiting a massive Kerr black hole. 
In order to obtain the precise theoretical gravitational waveform, 
we need to solve the self-force problem 
and problems in the second order perturbations~\cite{Nakano:2007hv}. 

In this paper, we focus on one aspect of the self-force problem, 
specifically to derive the retarded field of a point source. 
As a first step, we consider the Klein-Gordon equation 
in the Schwarzschild spacetime, but do not decompose it into spherical
harmonics, in order to model perturbations 
in the more generic Kerr background. 
Recently, Barack and Golbourn~\cite{Barack:2007jh} 
have discussed this equation in (2+1)-dimensions as 
derived by the mode decomposition in the azimuthal direction. 
Another treatment is proposed here to deal with this problem globally. 
We also note that there is a method to use a narrow Gaussian 
rather than a Dirac $\delta$-function~\cite{LopezAleman:2003ik}. 
It is, however,  difficult to ascertain the error introduced 
by smearing the particle and if this is accurate enough for self force
computations. 

%%%%%%%%%%%%%%%%%%%%%%%%%%%%%%%%%%%%%%%%%%%%%%%%%%%%%%%%%%%%%%%%%%%%%%
\section{Formulation} 
%%%%%%%%%%%%%%%%%%%%%%%%%%%%%%%%%%%%%%%%%%%%%%%%%%%%%%%%%%%%%%%%%%%%%%

When we calculate the $(2+1)$-dimensional ($D$) equation
derived from the $4D$ Klein-Gordon equation by the azimuthal mode decomposition, 
the resulting equation is not exactly same as the $(2+1)D$ wave equation. 
By transforming the scalar field, we can obtain an equation 
which includes the $(2+1)D$ d'Alambertian and a remainder. 
Then, we remove the $2D$ $\delta$-function 
in the source term. 

In order to do so, 
we consider the Schwarzschild metric in the isotropic coordinates, 
\begin{eqnarray}
ds^2 &=&
-(2\rho-M)^2/(2\rho+M)^2 dt^2+\left(1+M/(2\rho)\right)^{4} 
\left( d\rho^2
+\rho^2\left(d\theta^2+\sin^2\theta d\phi^2\right) \right) \,. 
\end{eqnarray}
This radial coordinate is related to that of the Schwarzschild, r, 
as $\rho = \left(r-M+\sqrt{r^2-2Mr}\right)/2$. 
In the above coordinates, the Klein-Gordon equation with 
a point source becomes 
\begin{eqnarray}
&& \Biggl[
-{\frac { \left( 2\,\rho+M \right) ^{2}}
{ \left( 2\,\rho-M \right) ^{2}}}\partial_t^{2}
+ {\frac {16 {\rho}^{4}}{ \left( 2\,\rho+M \right) ^{4}}}\partial_{\rho}^{2}
+{\frac {128{\rho}^{5}}
{ \left( 2\,\rho - M \right)  \left( 2\,\rho+M \right) ^{5}}}\partial_\rho
+{\frac {16 {\rho}^{2}}{ \left( 2\,\rho+M \right) ^{4}}}
\left(\partial_{\theta}^2 + \cot\theta \partial_{\theta} 
+\frac{1}{\sin^2 \theta}\partial_{\phi}^2 
\right)\Biggr] 
\nonumber \\ && \qquad \times \psi(t,\rho,\theta,\phi) 
= - q \int_{-\infty}^{\infty} d\tau 
\frac{64\rho^4\delta(t-t_z(\tau))\delta(\rho-\rho_z(\tau))\delta(\theta-\theta_z(\tau))
\delta(\phi-\phi_z(\tau))}
{(2\rho-M)(2\rho+M)^5 \sin\theta} 
 \,. 
\end{eqnarray}
Here, we use the azimuthal mode decomposition, 
$\psi(t,r,\theta,\phi)=\sum_{m=-\infty}^{\infty}
\psi_m(t,\rho,\theta) e^{im\phi}$, 
and transform the field $\psi_m$ as 
$\psi_m=2\sqrt{\rho/
[\left( 2\,\rho+M \right)  \left( 2\,\rho-M \right) \sin \theta ]}
\, \chi_m$. 
Then, $\chi_m$ satisfies the following equation. 
\begin{eqnarray}
&& 
\Biggl[
-\frac{1}{16}\frac{(2\rho+M)^6}{(2\rho-M)^2\rho^4}
\partial_t^2+\partial_{\rho}^2+\frac{1}{\rho} \partial_{\rho} 
+\frac{1}{\rho^2}\left( \partial_{\theta}^2 
-\frac{1}{\sin^2 \theta}
\left(m^2 - \frac{1}{4}\frac{(4\rho^2+M^2)^2-16\rho M \cos^2 \theta}
{(2\rho+M)^2(2\rho-M)^2} \right)
 \right) \Biggr] 
\nonumber \\ && \qquad \times \chi_m(t,\rho,\theta) 
= - q \int_{-\infty}^{\infty} d\tau 
\frac{2\,\delta(t-t_z(\tau))\delta(\rho-\rho_z(\tau))\delta(\theta-\theta_z(\tau))}
{\sqrt{(2\rho-M)(2\rho+M)\rho \sin\theta}} 
e^{-im \phi_z(\tau)} \,.
\label{eq:medWE}
\end{eqnarray}
Next, we use a new time coordinate which is defined by 
$T = \int^t dt \,4\,(2\rho_z(t)-M)\rho_z(t)^2/(2\rho_z(t)+M)^3$, 
where $\rho_z$ is obtained by solving a geodesic equation. 
Using this, we derive an equation which can be divided into 
the $(2+1)D$ d'Alambertian of the flat case $\Box^{(2+1)}$ and a remainder. 
\begin{eqnarray}
{\cal L}_m \, \chi_m(T,\rho,\theta) 
&=& \left(\Box^{(2+1)} + {\cal L}_m^{rem} \right) \chi_m(T,\rho,\theta) 
= S_m(T,\rho,\theta) \,;
\nonumber \\ 
\Box^{(2+1)} &=& -\partial_T^2+\partial_\rho^2 
+(1/\rho)\partial_\rho + (1/\rho^2) \partial_{\theta}^2 \,, 
\nonumber \\ 
{\cal L}_m^{rem} &=& 
\left(
1-\frac{(2\rho_z(T)-M)^2\rho_z(T)^4(2\rho+M)^6}{(2\rho_z(T)+M)^6(2\rho-M)^2\rho^4}
\right)\partial_T^2
\nonumber \\ && \qquad 
- \frac{2 (4\rho_z(T)-M) (2\rho_z(T)-M)(2\rho+M)^6 \rho_z(T)^3 M }
{ (2\rho_z(T)+M)^7(2\rho-M)^2\rho^4 }\left(\frac{d\rho_z(T)}{dT}\right)
\partial_T
\nonumber \\ && \qquad 
-\frac{1}{\rho^2 \sin^2 \theta}
\left(m^2 - \frac{1}{4}\frac{(4\rho^2+M^2)^2-16\rho M \cos^2 \theta}
{(2\rho+M)^2(2\rho-M)^2} \right) 
\,, 
\nonumber \\ 
S_m(T,\rho,\theta) &=& - q \int_{-\infty}^{\infty} d\tau 
\frac{2\,\delta(t(T)-t_z(\tau))\delta(\rho-\rho_z(\tau))\delta(\theta-\theta_z(\tau))}
{\sqrt{(2\rho-M)(2\rho+M)\rho \sin\theta}} 
e^{-im \phi_z(\tau)} \,. 
\label{eq:formal}
\end{eqnarray}

To remove the $\delta$-function in the source term, we set 
\begin{eqnarray}
\chi_m(T,r,\theta) &=& \chi_m^S(T,r,\theta) + \chi_m^{rem}(T,r,\theta) \,,
\end{eqnarray} 
where we define the new functions, $\chi_m^S$ and $\chi_m^{rem}$ as
calculated from 
\begin{eqnarray}
\Box^{(2+1)} \chi_m^S(T,r,\theta) = S_m(T,r,\theta) \,;
\quad 
{\cal L}_m \,\chi_m^{rem}(T,r,\theta) = -{\cal L}_m^{rem} \chi_m^S(T,r,\theta) 
= S_m^{(eff)}(T,\rho,\theta) \,.
\label{eq:divisionEq}
\end{eqnarray}
This decomposition of $\chi_m$ does not have any physical-meaning, 
i.e., $\chi_m^S$ is not identified as the singular part to be removed 
in the self-force calculation. 
Note that ${\cal L}_m^{rem}$ includes a second-order derivative. 
But, since the factor of $\partial_T^2$ is zero at the particle location, 
the singular behavior of the effective source $S_m^{(eff)}$ weakens. 
The derivation of the singular field $\chi_m^S$ 
can be performed through the Green function, 
\begin{eqnarray}
G(T,{\bf x};T',{\bf x'}) &=& 
\theta((T-T')-|{\bf x} - {\bf x'}|)
/\bigl(2\pi\sqrt{(T-T')^2-|{\bf x} - {\bf x'}|^2}\bigr)
 \,,
\end{eqnarray}
where $|{\bf x} - {\bf x'}|=(\rho^2+{\rho'}^2-2\,\rho\,{\rho'}\cos(\theta-\theta'))^{1/2}$ 
and $\chi_m^S$ is calculated by the following integral 
\begin{eqnarray}
\chi_m^S(T,\rho,\theta) &=& \int dT' \rho' d\rho' d\theta' \,
G(T,{\bf x};T',{\bf x'}) S_m(T',\rho',\theta') \,.
\label{eq:formalS}
\end{eqnarray} 

%%%%%%%%%%%%%%%%%%%%%%%%%%%%%%%%%%%%%%%%%%%%%%%%%%%%%%%%%%%%%%%%%%%%%%
\section{Circular Orbit Case} 
%%%%%%%%%%%%%%%%%%%%%%%%%%%%%%%%%%%%%%%%%%%%%%%%%%%%%%%%%%%%%%%%%%%%%%

We consider a particle in circular orbit, 
$z^{\alpha}(\tau) = \{u^t \tau,\,r_0,\,\pi/2,\,u^{\phi} \tau \}$, 
where $u^t = \sqrt{r_0/(r_0-3M)}$ and $u^{\phi} = \sqrt{M/[r_0^2(r_0-3M)]}$. 
Here, we focus on the $m \neq 0$ modes; the $m=0$ mode 
can be dealt with by the same treatment. 

%%%%%%%%%%%%%%%%%%%
\subsection{Singular field}
%%%%%%%%%%%%%%%%%%%

First, the relationship between the new time coordinate $T$ 
and the Schwarzschild time $t$ is obtained analytically as 
$
T = 4\,(2\rho_0-M)\rho_0^2/(2\rho_0+M)^3\, t
$ where 
$\rho_0 = 1/2(r_0-M+\sqrt{r_0^2-2Mr_0})$. 
Note that in general, for non circular orbits, we need a numerical integration to derive 
this relationship. From Eq.~(\ref{eq:formalS}), 
the singular field is derived as 
\begin{eqnarray}
\chi_m^S(t,r,\theta) 
%&=& 
%\frac{1}{2\pi} \int_{1}^{\infty} d{\cal T} 
%\frac{2\,\sqrt{\rho_0}}{u^t \sqrt{(2\rho_0+M) (2\rho_0-M)} } 
%e^{-im \Omega t}
%\frac{e^{im \Omega (2\rho_0+M)^3 |{\bf x} - {\bf x_z}|{\cal T}/ (4\,(2\rho_0-M)\rho_0^2) }}
%{\sqrt{{\cal T}^2-1}} 
%\nonumber \\ 
&=& \frac{i}{4}\frac{2\,\sqrt{\rho_0}}{u^t \sqrt{(2\rho_0+M) (2\rho_0-M)} } 
H_0^{(1)} 
\left(
\frac{(2\rho_0+M)^3}{4\,(2\rho_0-M)\rho_0^2} m \Omega |{\bf x} - {\bf x_z}|
\right) 
 e^{-im \Omega t} \,,
\end{eqnarray}
where 
%${\cal T}=4\,(2\rho_0-M)\rho_0^2/(2\rho_0+M)^3
%\,[(t-T_0)/|{\bf x} - {\bf x_z}|]$, 
$\Omega = u^\phi/u^t$ 
and the spatial difference is given by 
$|{\bf x} - {\bf x_z}|=\sqrt{\rho^2+\rho_0^2-2\,\rho\, \rho_0 \sin \theta}$ 
and $H_0^{(1)}$ is the Hankel function. 
The local behavior of the above solution 
near the particle location is obtained as 
\begin{eqnarray}
\chi_m^S(t,r,\theta) \sim \chi_m^{SL}(t,r,\theta) = - \frac{1}{2\pi}
\frac{2\,\sqrt{\rho_0}}{u^t \sqrt{(2\rho_0+M) (2\rho_0-M)} } 
\ln 
\left(
\frac{(2\rho_0+M)^3}{4\,(2\rho_0-M)\rho_0^2} m \Omega |{\bf x} - {\bf x_z}|
\right)
e^{-im \Omega t}  \,.
\end{eqnarray}
Therefore, $S_m^{(eff)}(t,\rho,\theta)$ 
shown by the dashed green curve in Fig.~\ref{fig:combm}, 
is singular at the particle location. 
In order to perform the numerical integration with higher accuracy, it is convenient 
to regularize the source term to be at least $C^0$ at the particle location. 

%%%%%%%%%%%%%%%%%%%
\subsection{Local behavior}
%%%%%%%%%%%%%%%%%%%

When we write the singular field as 
$\chi_m^S = \chi_m^{SL} + {\hat \chi}_m^S$, 
${\hat \chi}_m^S$ is finite at the particle location. 
Then, the effective source in Eq.~(\ref{eq:divisionEq}) becomes 
\begin{eqnarray}
S_m^{(eff)}(t,\rho,\theta) &=& - \frac{1}{2\pi}
\frac{2\,\sqrt{\rho_0}}{u^t \sqrt{(2\rho_0+M) (2\rho_0-M)} } 
\ln 
\left(
\frac{(2\rho_0+M)^3}{4\,(2\rho_0-M)\rho_0^2} m \Omega |{\bf x} - {\bf x_z}|
\right)
e^{-im \Omega t}
\nonumber \\ && \times 
\Biggl(
-\frac{1}{\rho^2 \sin^2 \theta}
\left(m^2 - \frac{1}{4}\frac{(4\rho^2+M^2)^2-16\rho M \cos^2 \theta}
{(2\rho+M)^2(2\rho-M)^2} \right) 
\nonumber \\ && \qquad 
-(m \Omega)^2\left(
\frac{(2\rho_0+M)^6}{(2\rho_0-M)^2\rho_0^4}
-\frac{(2\rho+M)^6}{(2\rho-M)^2\rho^4}
\right)
\Biggr)
- {\cal L}_m^{rem} {\hat \chi}_m^S(t,r,\theta)
\,,
\label{eq:cir_Smeff}
\end{eqnarray}
Note that the third line 
of the R.H.S. is at least $C^0$ at the location of the particle. 

To remove the logarithmic divergence in the source, 
we introduce 
\begin{eqnarray}
\chi_m^{rem,S}(t,\rho,\theta) &=& 
-\frac{1}{16}\, |{\bf x} - {\bf x_z}|^2 
\ln 
\left(
\frac{(2\rho_0+M)^3}{4\,(2\rho_0-M)\rho_0^2} m \Omega |{\bf x} - {\bf x_z}|
\right)
{{\rho_0}}^{19/2} \left( 2\,\rho-M \right) ^{3} {e^{-im\Omega\,t}} 
\nonumber \\ && 
\times 
\frac{
64\,{m}^{2}{\rho}^{4}-32\,{m}^{2}{\rho}^{2}{M}^{2}
+4\,{m}^{2}{M}^{4}+16\, \cos^2 \theta 
{\rho}^{2}{M}^{2}
%%-{M}^{4}-8\,{\rho}^{2}{M}^{2}-16\,{\rho}^{4} 
-(4\rho^2+M^2)^2}
{
{\pi } u^t 
\left( 2\,{\rho_0}+M \right) ^{5/2} \left( 2\,{\rho_0}-M \right) ^{11/2} 
{\rho}^{11}\sin^2 \theta  }
\,.
\label{eq:remS}
\end{eqnarray}
Using this regularization function, we obtain a source 
$S_m^{reg,I}$ for the function $\chi_m^{rem}-\chi_m^{rem,S}$ as 
\begin{eqnarray}
S_m^{reg,I}(t,\rho,\theta) &=& S_m^{(eff)}(t,\rho,\theta) 
- {\cal L}_m \chi_m^{rem,S}(t,\rho,\theta) \,.
\end{eqnarray}
The local behavior of $S_m^{reg,I}$, 
which is shown by the solid green curve 
in Fig.~\ref{fig:combm} (b), 
is an "$x \ln |x|$ for $x \rightarrow 0$" type, i.e., 
$C^0$ around the particle location. 

%%%%%%%%%%%%%%%%%%%
\subsection{Boundary behaviors} 
%%%%%%%%%%%%%%%%%%%

We now focus on the behaviors of the source term at the two boundaries, i.e 
at the horizon of the large hole and spatial infinity. 
The source for a final regularized function $\chi_m^{reg}$ 
must go like $O(\rho^{-2})$ for $\rho \rightarrow \infty$ 
in the case of the $m=0$ mode 
and $O(\rho^{-3/2})$ for the $m \neq 0$ mode because of integrability conditions. 
More precisely, the source for the regularized function of $\psi_m$ 
derived by numerical calculations has a factor $\sim 1/\rho^{1/2}$. 
For $\rho \rightarrow M/2$, the source should be zero, 
i.e., the behavior should be a power of $(\rho-M/2)$ greater than $1/2$.
To regularize the source at the boundaries, 
we note that the source contribution from $\chi_m^{rem,S}$ is well behaved. 
This means that the ill behaviors of the source 
arise from $\chi_m^S$. 
Therefore, it is convenient to use asymptotic behaviors 
of $\chi_m^S$ (and some correction factor) for regularization. 

\begin{figure}[ht]
\center
\epsfxsize=4.6cm
\epsfbox{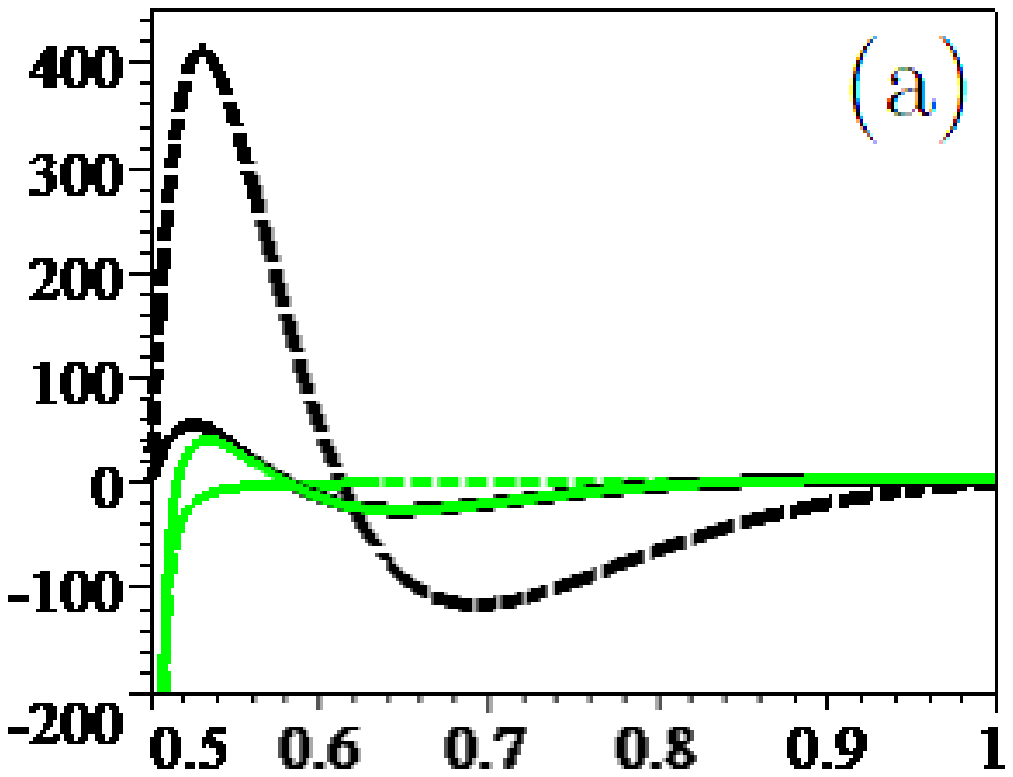}
\hspace{5mm}
\epsfxsize=4.7cm
\epsfbox{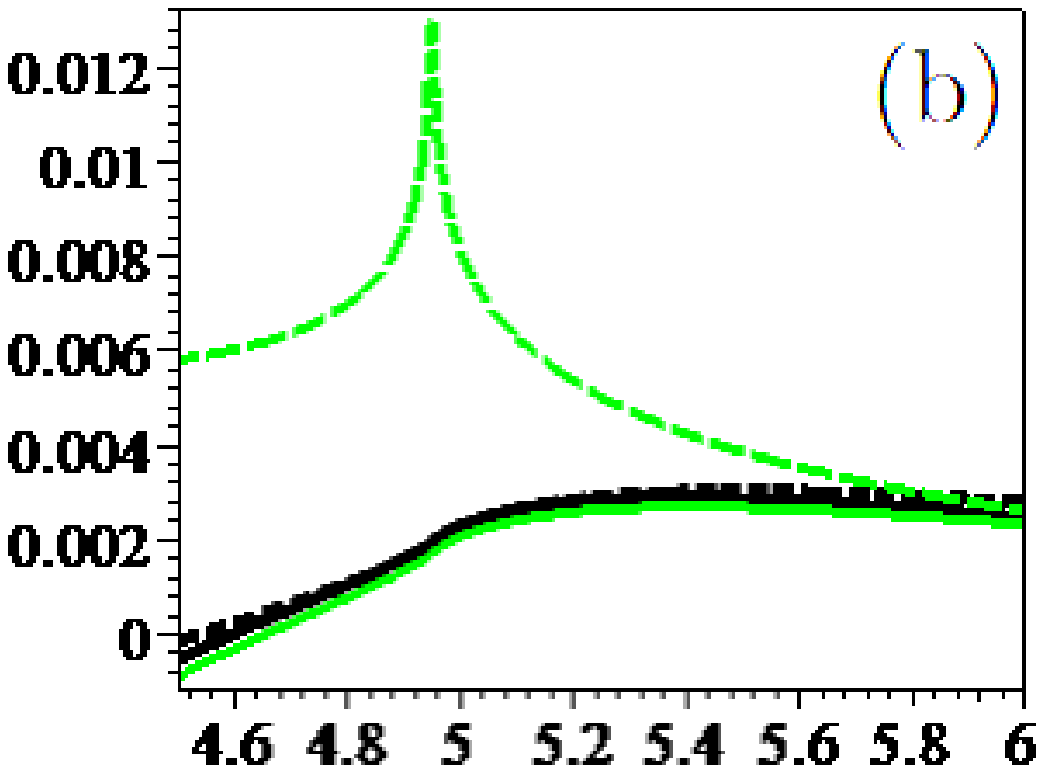}
\hspace{5mm}
\epsfxsize=5.0cm
\epsfbox{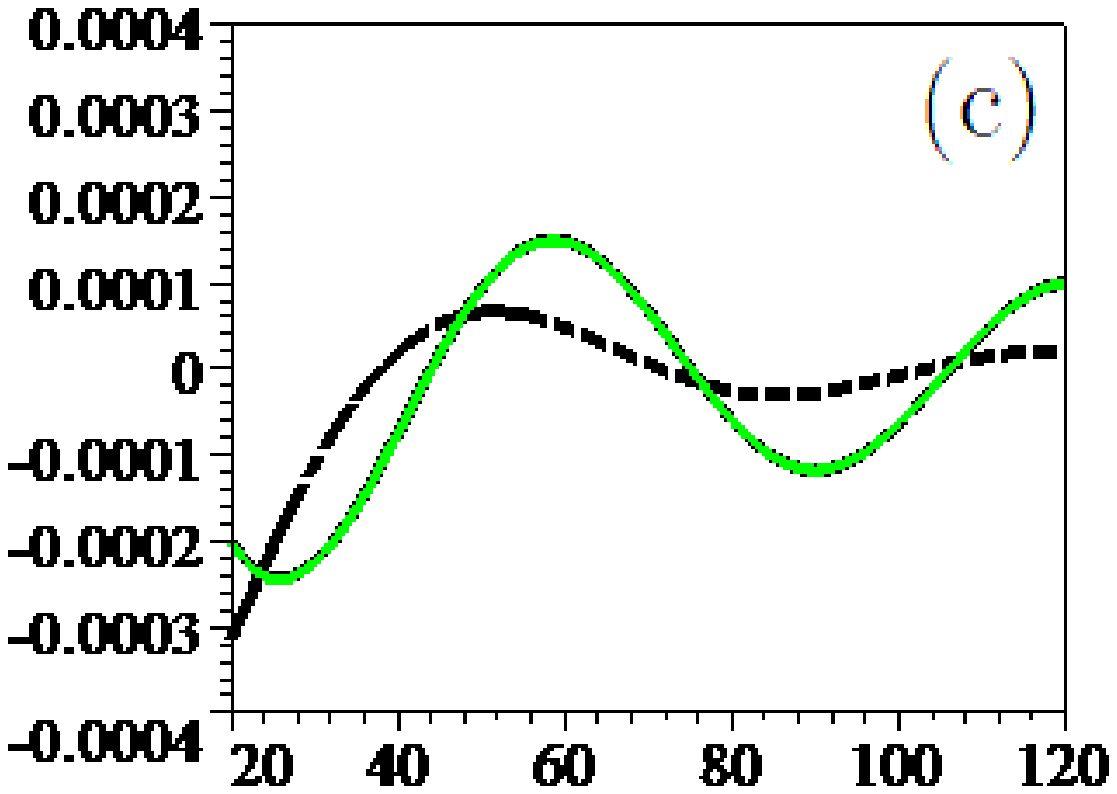}
\caption{
Plot for the $m=1$ mode of $S_m$ with respect to $\rho$. 
$S_1^{(eff)}$, $S_1^{reg,I}$, $S_1^{reg,h}$ and $S_1^{reg,f}$ 
are shown by the dashed green, solid green, 
solid black and dashed black curve, respectively.
}
\label{fig:combm}
\end{figure}

First, for the regularization near the horizon, 
we use the regularization function $\chi_m^h$ 
which is too long to be shown here. Then, the source 
for the function $\chi_m^{rem}-\chi_m^{rem,S}-\chi_m^h$ becomes 
\begin{eqnarray}
S_m^{reg,h}(t,\rho,\theta) &=& S_m^{reg,I}(t,\rho,\theta) 
- {\cal L}_m \chi_m^{h}(t,\rho,\theta) 
\,,
\end{eqnarray}
This $S_m^{reg,h}$ is shown by the solid black curve in Fig.~\ref{fig:combm} 
and behaves as $O(\rho^{-1/2})$ for large $\rho$. 
To regularize it, 
we use the regularization function, 
\begin{eqnarray}
\chi_m^{\infty}(t,\rho,\theta) &=& 
 - \sqrt{2}\,i\,\frac{\rho_0^{3/2}}{(2\,\rho_0 + M)^{2}\,u^t\,\sqrt{\pi }\,\sqrt{m\,\Omega \,\rho }\,\rho ^{7}} 
\,e^{( - i\,m\,\Omega \,t)}\,\left(\rho  - {\displaystyle \frac {M}{2}} \right)^{3} 
(\rho ^{2} + \rho_0^{2} - 2\,\rho_0\,\rho 
\,\sin \theta)^{2} 
\nonumber \\ &&
\times 
\,\exp
\left(  \!  \frac{1}{4}i\,\frac {(2\,\rho_0 + M)^{3}\,m\,\Omega \,
\sqrt{\rho ^{2} + \rho_0^{2} - 2\,\rho_0\,\rho 
\,\sin \theta}}{(2\,\rho_0 - M)\,\rho_0^{2}} 
- \frac {\pi }{4}i \!  \right) 
\,. 
\end{eqnarray}
The final source for the regularized function 
$\chi_m^{reg}=\chi_m^{rem}-\chi_m^{rem,S}-\chi_m^h-\chi_m^{\infty}$ becomes 
\begin{eqnarray}
S_m^{reg,f}(t,\rho,\theta) &=& S_m^{reg,h}(t,\rho,\theta) 
- {\cal L}_m \chi_m^{\infty}(t,\rho,\theta) 
\,,
\end{eqnarray}
which is used in the numerical calculation.  
This $S_m^{reg,f}$ is shown by the dashed black curve in Fig.~\ref{fig:combm}, 
and behaves like $O(\rho^{-3/2})\, \times$ 
(an oscillation factor with respect to $\rho$) for large $\rho$. 

%%%%%%%%%%%%%%%%%%%%%%%%%%%%%%%%%%%%%%%%%%%%%%%%%%%%%%%%%%%%%%%%%%%%%%
\section{Discussion} 
%%%%%%%%%%%%%%%%%%%%%%%%%%%%%%%%%%%%%%%%%%%%%%%%%%%%%%%%%%%%%%%%%%%%%%

In this paper, we obtained the regularized effective source 
which is $C^0$ at the location of the particle, 
and $O(\rho-M/2)$ near the horizon. 
The behavior at infinity is $O(\rho^{-3/2}) \, \times $ 
(an oscillation factor with respect to $\rho$) which allows
straightforward numerical integration.

When we consider the extension of this formulation 
to the Kerr background case, we can also extract
a similar differential operator 
to that of Eq.~(\ref{eq:formal}). 
In the case of gravitational perturbations, 
we have ten field equations for the linear perturbation 
in the Lorenz gauge. (See \cite{Barack:2005nr}.)
The same treatment discussed in this paper 
is applicable to those equations.

%===========================%
\subsection*{Acknowledgments}
%===========================%

We would like to thank N.~Sago and  H.~Tagoshi for useful discussions.

%%%%%%%%%%%%%%%%%%%%%%%%%%%%%%%%%%%%%%%%%%%%%%%%%%%%%%%%%%%%%%%%%%%%%%
%%%%%%%%%%%%%%%%%%%%%%%%%%%%%%%%%%%%%%%%%%%%%%%%%%%%%%%%%%%%%%%%%%%%%%

\end{document}